\documentclass{elsart}

\usepackage{amssymb}

\begin{document}

\begin{frontmatter}



\title{Conductivity of thermally fluctuating superconductors in two dimensions}


\author{Subir Sachdev}
\ead[url]{http://pantheon.yale.edu/\~\/subir}
\address{Department
of Physics, Yale University,\\ P.O. Box 208120, New Haven CT
06520-8120, USA}

\begin{abstract}
We review recent work on a continuum, classical theory of thermal
fluctuations in two dimensional superconductors. A functional
integral over a Ginzburg-Landau free energy describes the
amplitude and phase fluctuations responsible for the crossover
from Gaussian fluctuations of the superconducting order at high
temperatures, to the vortex physics of the Kosterlitz-Thouless
transition at lower temperatures. Results on the structure of this
crossover are presented, including new results for corrections to
the Aslamazov-Larkin fluctuation conductivity.
\end{abstract}

\end{frontmatter}

% main text
\section{Introduction}
\label{sec:intro}

The subject of fluctuation conductivity of superconductors is an
old one, and many classic results can be found in a recent review
article by Larkin and Varlamov \cite{larkin}. Essentially all of
this work is restricted to the first corrections to the BCS
theory, and involving Gaussian fluctuations in the Cooper pair
propagator. Such an approach is valid provided the Ginzburg
fluctuation parameter is small.

In two dimensions, we know that as the superconducting critical
temperature ($T_c$) is approached from above, there is ultimately
a crossover from such a Gaussian fluctuation regime to the vortex
physics of the Kosterlitz-Thouless (KT) transition. While
universal characteristics of the conductivity in the latter regime
are well understood \cite{hn}, there is little quantitative
understanding of precisely how such a crossover occurs. In the
context of the small coherence length cuprate superconductors,
where there is a wide regime of temperatures where the Ginzburg
fluctuation parameter is large, a quantitative analysis of such a
regime is surely important in deciphering the importance of
pairing fluctuations.

This paper will address the Gaussian-to-vortex crossover in the
fluctuation conductivity using an approach suggested in
Refs.~\cite{ssrelax,nature}; the same approach will also be
extended to describe the spin-wave-to-vortex crossover at low
temperatures. Using the proximity of the cuprates to a
superconductor-insulator quantum transition, it was argued that
the fluctuation regime could be described by a universal continuum
limit of the Ginzburg-Landau free energy. This is described by the
following partition function for the complex pairing order
parameter $\Psi ({\bf r})$:
\begin{eqnarray}
\mathcal{Z}_{GL} &=& \int \mathcal{D} \Psi ({\bf r})
e^{-\mathcal{F}_{GL}/(k_B T)} \label{gl} \\
 \mathcal{F}_{GL} &=& \int
d^2 r \left[ \frac{\hbar^2 |\nabla_{{\bf r}} \Psi ({\bf r})|^2}{2
m^{\ast}} + a(T) |\Psi ({\bf r})|^2 + \frac{b}{2} |\Psi ({\bf r})
|^4 \right] \nonumber
\end{eqnarray}
Here $m^{\ast}$, $a(T)$ and $b$ are parameters dependent upon the
physics of the underlying electrons which have been integrated
out; the function $a(T)$ vanishes at the mean-field transition
temperature $a(T_c^{MF})=0$. Note that the previous work
\cite{larkin} was at the level of a mean field treatment of
$\mathcal{F}_{GL}$, or its one loop fluctuations. Furthermore, the
strong fluctuations near the KT transition have been invariably
treated in previous work in terms of theory for the vortices alone
or in terms of a phase-only XY model. Refs.~\cite{ssrelax,nature}
proposed that the combined amplitude and phase fluctuations
implied by (\ref{gl}) should be taken seriously, from temperatures
well above $T_c$, and across the vortex physics of the KT
transition. This paper will present a few results on the
electrical conductivity associated with the continuum theory of
equal-time fluctuations described by (\ref{gl}). Our results for
the structure of the crossover contains non-perturbative
information in the form of a few universal numerical constants
which were computed in recent computer simulations
\cite{ssrelax,nikolay1,nikolay2}.

\section{Review of static theory}
\label{sec:static}

This section reviews a few important static properties of
$\mathcal{Z}_{GL}$ \cite{ssrelax,demler}.

A key first point is that the theory (\ref{gl}) cannot predict
actual $T_c$ on its own, as its value is dependent upon a short
distance cutoff. We treat $T_c$ as an input parameter, and then
find that all subsequent cutoff dependence can be safely
neglected. Knowing $T_c$, it is useful to define the dimensionless
coupling $g$
\begin{equation}
g \equiv \frac{\hbar^2}{m^{\ast} b} \left[ \frac{a(T)}{k_B T} -
\frac{a(T_c)}{k_B T_c} \right], \label{defg}
\end{equation}
which should be a monotonically increasing function of $T$
crossing zero at exactly $T=T_c$. If we take $a(T)$ to be a simple
linearly increasing function of temperature ($a(T) = a_0
(T-T_c^{MF})$), then
\begin{equation}
g = \frac{\rho_s (0)}{k_B T_c} \left( 1 - \frac{T_c}{T} \right),
\label{rhos}
\end{equation}
where $\rho_s (T)$ is the helicity modulus\footnote{Of course, we
do not expect the classical theory in (\protect\ref{gl}) to be
valid as $T \rightarrow 0$ because quantum effects will become
important; by $\rho_s (0) \equiv -\hbar^2 a(0) /(m^{\ast} b)$ we
mean the energy obtained by extrapolating the present
approximation for $a(T)$ to $T=0$.} of $\mathcal{Z}_{GL}$ (the
London penetration depth is $\lambda_L^{-2} = 4 \pi e^{\ast 2}
\rho_s (T)/(\hbar^2 c^2)$, where $e^{\ast} = 2e$, the Cooper pair
charge). In this framework, the values of $\rho_s (0)$ and $T_c$
are the only two needed inputs to predict most physical properties
of $\mathcal{Z}_{GL}$ (including the full $T$ dependence of
$\rho_s (T)$). Note that $\rho_s (0)$ and $T_c$ are, in general,
independent of each other, and the Nelson-Kosterlitz relation
\cite{nk} only constraints $\rho_s (T_c)/T_c = 2/\pi$.

It was shown in Ref.~\cite{ssrelax,demler} that the structure of
expansions away from simple limiting mean-field regimes of
$\mathcal{Z}_{GL}$ are best understood in terms of two new
dimensionless couplings, $\mathcal{G}$ and $\mathcal{G}_D$. These
couplings are most useful in high and low $T$ limits,
respectively.

For $T \geq T_c$, we work with the coupling $\mathcal{G}>0$ given
by
\begin{equation}
g = \frac{6}{\mathcal{G}} + \frac{1}{\pi} \ln \left(
\frac{\mathcal{C} T}{\mathcal{G}T_c} \right), \label{defgG}
\end{equation}
where $\mathcal{C}$ is a universal numerical constant. Its value
is determined by the condition that at the critical point with
$g=0$, $\mathcal{G}=\mathcal{G}_c = 96.9 \pm 3$ a universal
critical value that has been determined in numerical simulations
\cite{ssrelax,nikolay1,nikolay2}. Using these results, we
obtain\footnote{$\mathcal{C}$ is related to the number $\xi_{\mu}$
computed in Refs.~\protect\cite{nikolay1,nikolay2} by $\mathcal{C}
= 6 \xi_{\mu}$.}
\begin{equation}
\mathcal{C} = 79.8 \pm 2.4 .\label{defC}
\end{equation}
A key fact behind our main results here is that for large $g$
(when $\mathcal{G}$ is small), all physical properties of
$\mathcal{Z}_{GL}$ can be expanded in a series that involves {\em
only} integer powers of $\mathcal{G}$. Re-expressing any such
series in terms of $1/g$ by (\ref{defgG}) then implies a number of
logarithmic terms in $g$: the coefficients and arguments of such
logarithms are therefore entirely specified by (\ref{defgG}). We
should also note here that, as discussed in Ref.~\cite{demler},
the combination of (\ref{defgG}) and (\ref{rhos}) implies that the
continuum theory $\mathcal{Z}_{GL}$ breaks down at a $T$ so large
($T > \pi \rho_s (0)/k_B$ for the $T$ dependence assumed in
(\ref{rhos})) that $\mathcal{G}$ ceases to be a monotonically
decreasing function of increasing $T$.

The equation (\ref{defgG}) also defines $\mathcal{G}$ for $T <
T_c$, but its value becomes exponentially large as $T \rightarrow
0$. Instead, for $T \leq T_c$, the useful coupling turns out to be
$\mathcal{G}_D>0$ defined by
\begin{equation}
g = - \frac{3}{\mathcal{G}_D} +  \frac{1}{\pi} \ln \left(
\frac{\mathcal{C} T}{\mathcal{G}_D T_c} \right)\label{defgd}
\end{equation}
The coupling $\mathcal{G}_D$ vanishes linearly as $T \rightarrow
0$, and is an increasing function of $T$. At low $T$ we can expand
physical properties of $\mathcal{Z}_{GL}$ in $\mathcal{G}_D$, and
again, this series involves {\em only} integer powers of
$\mathcal{G}_D$. The relationship (\ref{defgd}) then implies
logarithms in the expansion in $1/|g|$.

\section{Conductivity}

A description of electrical transport requires an extension of the
partition function $\mathcal{Z}_{GL}$ to a dynamic theory. We do
this in the simplest `model A' theory, in which the time evolution
of $\Psi ({\bf r}, t)$ is given by
\begin{equation}
\tau \frac{\partial \Psi ({\bf r}, t)}{\partial t} = -
\frac{\delta \mathcal{F}_{GL}}{\delta \Psi^{\ast} ({\bf r}, t)} +
\zeta ({\bf r}, t) \label{langevin}
\end{equation}
where $\tau$ is a damping parameter which has no singularities at
$T=T_c$, and $\zeta$ is a Gaussian random noise source with the
correlator
\begin{equation}
\left\langle \zeta^{\ast} ({\bf r}^{\prime}, t^{\prime}) \zeta
({\bf r}, t) \right\rangle = 2 \tau k_B T \delta\left({\bf r} -
{\bf r}^{\prime}\right) \delta\left(t-t^{\prime}\right)
\label{noise}
\end{equation}
Wickham and Dorsey \cite{dorsey} have already computed the
conductivity in the theory (\ref{gl}, \ref{langevin}, \ref{noise})
as an expansion in powers of $b$, and many needed results can be
obtained simply by transforming their results to spatial dimension
$d=2$ (they performed an analysis of perturbation theory in powers
of $(4-d)$ for different purposes). As the steps in the
calculation are essentially identical to theirs, no details of the
will be presented here.

Our results are most easily expressed in terms of a dimensionless
measure of the damping parameter, which we define as
\begin{equation}
\overline{\tau} \equiv \frac{\hbar \tau}{m^{\ast} b}
\label{damping}
\end{equation}
In BCS theory, the damping arises from decay of Cooper pairs into
electrons; assuming a linear dependence of $a(T)$ on $T$ as in
(\ref{rhos}), BCS theory predicts \cite{larkin}
\begin{equation}
\overline{\tau}_{BCS} = \frac{\pi\rho_s (0)}{8k_B T_c}.
\label{taubcs}
\end{equation} Of course, as $T \rightarrow 0$, the gapless
quasiparticle excitations disappear, and then a simple damping
term as in (\ref{langevin}) is not expected to be valid. Our low
$T$ results for $\mathcal{Z}_{GL}$ should therefore be treated
with caution.

The main new claim of this paper is that the frequency ($\omega$)
dependent conductivity can be written as
\begin{equation}
\sigma (\omega ) = \frac{e^{\ast 2}\overline{\tau} }{\hbar} ~
\Upsilon \left( \mathcal{G}, \frac{\hbar \omega \mathcal{G}
\overline{\tau} }{12 k_B T} \right) \label{upsilon}
\end{equation}
where $\Upsilon$ is a universal function (the factor of 12 in the
second argument is for future notational convenience). The result
(\ref{upsilon}) applies for all values of $g$ and $\mathcal{G}$,
but for $g \ll 1$, a different scaling form involving
$\mathcal{G}_D$ is more convenient (see Section~\ref{gsmall}). We
discuss the structure of this function for the different limiting
ranges of $g$ in the subsections below.

\subsection{$ g \gg 1$}
\label{glarge}

At $T\geq T_c$, where $g \gg 1$, there is an expansion for
$\Upsilon$ in {\em integer powers of its first argument
$\mathcal{G}$}. The results of Wickham and Dorsey \cite{dorsey},
when mapped to $d=2$, immediately yield an expansion of $\Upsilon$
to order $\mathcal{G}^3$. For the d.c. conductivity, these results
are
\begin{equation}
\Upsilon (\mathcal{G}, 0) = \frac{\mathcal{G}}{48\pi} +
\mathcal{A} \mathcal{G}^3 + \mathcal{O} (\mathcal{G}^4)
\label{series}
\end{equation}
The first term in (\ref{series}) is precisely the old result of
Aslamazov and Larkin \cite{larkin}: this can be checked by noting
that for $g \gg 1$, $\mathcal{G} \approx 6/g$ from (\ref{defgG}),
and by using (\ref{rhos}) and the value for
$\overline{\tau}_{BCS}$ in (\ref{taubcs}), which yields ultimately
$\sigma = e^{\ast 2} T/(64 \hbar (T-T_c))$. The numerical constant
$\mathcal{A}$ involves a number of different contributions, and
can be written as
\begin{eqnarray}
\mathcal{A} &=& \frac{J}{216 \pi} + \frac{2}{9} \left(
\widetilde{I}_2^a (0) + \widetilde{I}_2^b (0) \right)
\nonumber \\
&~&~~~~~~~~~~~~- \frac{32}{27} \left( 4 \widetilde{I}_b (0) +
\widetilde{I}_c^{(1)} (0) + \widetilde{I}_c^{(2)} (0) \right).
\label{aval}
\end{eqnarray}
The first term in (\ref{aval}) is a `mass' renormalization'
correction to the leading order term in (\ref{series}), and arises
from (2.1) of Ref.~\cite{ssrelax} with $J=0.014842966\ldots$; the
remaining terms involving the $\widetilde{I}$'s  are associated
with 3-loop diagrams evaluated by Wickham and Dorsey, and are
specified in Eqs. (4.26), (4.28), (5.2), (5.3), and (5.4)
respectively of Ref.~\cite{dorsey}. The momentum integrals in all
the $\widetilde{I}$'s have to be evaluated in $d=2$. This was done
numerically, and we obtained
\begin{eqnarray}\widetilde{I}_2^a (0)
&=& 4.17384 \times 10^{-6} \nonumber \\
\widetilde{I}_2^b (0) &=& -6.92307 \times 10^{-6} \nonumber \\
\widetilde{I}_b (0) &=&-1.06867 \times 10^{-7} \nonumber \\
\widetilde{I}_c^{(1)} (0) &=&-6.41276 \times 10^{-7} \nonumber
\\
\widetilde{I}_c^{(2)} (0) &=&-2.93446 \times 10^{-7} \nonumber \\
\mathcal{A} &=& 2.28769 \times 10^{-5}
\end{eqnarray}

We can combine (\ref{defgG}), (\ref{upsilon}) and (\ref{series})
to obtain the following corrections to the Aslamazov-Larkin result
valid for $g \gg 1$:
\begin{eqnarray}
\sigma (0) &=& \frac{e^{\ast 2} \overline{\tau}}{8 \pi \hbar g}
\left[ 1 + \frac{1}{\pi g} \ln (13.3 g T/T_c)
\right. \nonumber \\
&~&~~~+  \frac{1}{\pi^2 g^2} \left( \ln^2 (13.3 g T/T_c)- \ln
(13.3gT/T_c) \right) \nonumber \\
&~&~~~ \left. + \frac{1728 \pi \mathcal{A}}{g^2} + \mathcal{O}
(1/g^3) \right] \label{key}
\end{eqnarray}
The overall pre-factor is the Aslamazov-Larkin result. The series
is (\ref{key}) one of the main new results of this paper. Using
the value of $g$ in (\ref{rhos}) as input, it yields the $T$
dependence of the fluctuation conductivity. Of course, it is more
accurate to numerically solve (\ref{defgG}) for $\mathcal{G}$ and
to then use (\ref{series}): such a procedure would account for all
logarithms.

The results for the frequency dependence of the conductivity are
much more cumbersome, and we quote only the terms to order
$\mathcal{G}^2$:
\begin{equation}
\Upsilon (\mathcal{G}, y) = \frac{\mathcal{G}}{48\pi} \frac{2i(-y
+ (i+y) \ln(1 - iy))}{y^2} + \mathcal{O} (\mathcal{G}^3).
\end{equation}
This result is obtained by taking the $d \rightarrow 2$ limit of
Eq. (3.7) in Ref.~\cite{dorsey}.

\subsection{$g \approx 0$} \label{gcrit}

The Kosterlitz-Thouless transition occurs at $\mathcal{G} =
\mathcal{G}_c$, and the system continues to be described by
(\ref{upsilon}) across the transition. The singularities of the
transition appear as singularities in  the function $\Upsilon$ as
a function of the $\mathcal{G}$. The functional form of these
singularities will similar to those in Ref.~\cite{hn} as a
function of $T$. Of course, there are no arbitrary scale factors
in the function $\Upsilon$ near this singularity, but these can
only be determination by suitable simulations.

\subsection{$ g \ll -1 $}
\label{gsmall}

 For $T<T_c$, it is useful to expand about a saddle point with
 $\Psi \neq 0$, and to organize the perturbation theory in powers
 of $\mathcal{G}_D$ defined in (\ref{defgd}). We therefore rewrite
 (\ref{upsilon}) as
 \begin{equation}
 \sigma (\omega ) = \frac{e^{\ast 2}}{\hbar^2} \rho_s (T) \delta
 (\omega) + \frac{e^{\ast 2}\overline{\tau} }{\hbar} ~
\Upsilon_D \left( \mathcal{G}_D, \frac{\hbar \omega \mathcal{G}_D
\overline{\tau} }{12 k_B T} \right) \label{upsilond}
\end{equation}
Again, both $\rho_s (T)$ and $\Upsilon_D$ can be expanded in a
series involving only integer powers of $\mathcal{G}_D$. For
$\rho_s (T)$, terms in this series were already computed in
Ref.~\cite{ssrelax}:
\begin{equation}
\frac{\rho_s (T)}{k_B T} = \frac{3}{\mathcal{G}_D} -
\frac{\mathcal{G}_D}{36} + \mathcal{O} (\mathcal{G}_D^2)
\end{equation}
We computed $\Upsilon_D$ to one loop order by the methods of
Ref.~\cite{dorsey}, and found
\begin{equation}
\Upsilon_D (\mathcal{G}_D, y ) = -\frac{\mathcal{G}_D}{12\pi}
\frac{\ln(1/2 - iy)}{(1+ 2 i y)} + \mathcal{O} (\mathcal{G}_D^2).
\end{equation}

\section{Conclusions}

This paper has obtained systematic corrections to the
Aslamazov-Larkin fluctuation conductivity of thermally fluctuating
superconductors in two dimensions. The main results are contained
in (\ref{upsilon}) and (\ref{series}), and take the form of an
expansion in integer powers of a renormalized dimensionless
coupling $\mathcal{G}$. When $\mathcal{G}$ is re-expressed in
terms of bare microscopic parameters (such as the temperature),
numerous logarithms appear. However, this microscopic relationship
is known exactly, and consequently, so is the structure of the all
the logarithms; this relationship involves non-perturbative
information in the form of universal constants which were computed
in recent numerical simulations \cite{ssrelax,nikolay1,nikolay2}.
The renormalized coupling $\mathcal{G}$ is related to a bare
coupling $g$ by (\ref{defgG}), and an expansion for the
conductivity in inverse powers of $g$ is in (\ref{key}). The
coupling $g$ is related to parameters in the Ginzburg-Landau free
energy by (\ref{defg}). Alternatively, assuming a linear
dependence of $a(T)$ on $T$, we can use the simpler expression
(\ref{rhos}), where $\rho_s (0) \equiv -\hbar^2 a(0) /(m^{\ast}
b)$, is the helicity modulus of the Ginzburg-Landau free energy
extrapolated to $T=0$.

It is useful to mention here the expected values of the
dimensionless parameter $\rho_s (0)/(k_B T_c)$ appearing in
(\ref{rhos}). In BCS theory, we can use standard expressions for
the parameters in (\ref{gl}), obtained for electrons in a simple
parabolic band \cite{larkin}: keeping only the linear $T$
dependence near $T_c$, and extrapolating to $T=0$ we obtain
\begin{equation}
\left. \frac{\rho_s (0)}{k_B T_c} \right|_{BCS} = \frac{E_F}{\pi
k_B T_c}
\end{equation}
where $E_F$ is the Fermi energy. The right-hand-side involves the
very large ratio $E_F/T_c$, and this ensures a large regime where
$g \gg 1$ above $T_c$. We can also compare to the numerical
studies of the cuprate superconductors in Ref~\cite{huse2}. They
defined a dimensionless parameter $\eta$, in terms of which we can
write $\rho_s (0)/(k_B T_c) = (2/\eta) (T_c^{MF}/T_c)$; for their
parameter values we have $\rho_s (0)/(k_B T_c) = 6.8$, which also
ensures a reasonable large regime over which the $g \gg 1$
expansion is useful \cite{demler}. Finally, we note that near a
superfluid-insulator quantum phase transition, $\rho_s (0)/(k_B
T_c)$ is the ratio of two energy scales which vanish at the
quantum critical point, and so is a universal number
\cite{ssrelax}: this universal number has been computed in a
simple model for such a transition.

It would be interesting to extend the present results to other
transport co-efficients, such as the Nernst effect
\cite{huse2,huse1}. The present quantitative approach is likely to
be most useful for observables \cite{demler} that do not diverge
at $T_c$.

%I thank Alan Dorsey for providing additional details of the
%calculations in Ref.~\cite{dorsey}, which aided in the
%determination of the $\widetilde{I}$'s in (\ref{aval}).
This research was supported by US NSF Grant DMR-0098226.

% \appendix

% \section{}
% \label{}

\end{document}